\documentclass[twocolumn,showpacs,preprintnumbers,amsmath,amssymb]{revtex4}
\usepackage{graphicx}
\usepackage{dcolumn}
\usepackage{bm}
\begin{document}

\title
{
Quantum Persistence: A Random Walk Scenario                                                                                    }                                                  
\author{Sanchari Goswami,  Parongama Sen}
\affiliation{
Department of Physics, University of Calcutta,92 Acharya Prafulla Chandra Road,
Calcutta 700009, India.
}
\author{Arnab Das}
\affiliation{
The Abdus Salam International Centre for Theoretical Physics,
Condense Matter and Statistical Physics Section,
Strada Costiera 11,
Trieste
34151,
Italy
}

\begin{abstract}
In this paper we extend the concept of persistence, well defined for classical
stochastic dynamics, to the context of quantum dynamics. We demonstrate
the idea via quantum random walk and a successive measurement scheme,
where persistence is defined as the time
during which a given site remains unvisited by the walker. We also
investigated the behavior of  related  quantities, e.g., the first-passage time
and the succession probability (newly defined),
etc. The study reveals power law  scaling  behavior of these
quantities with new exponents. 
Comparable features of the   
classical and the quantum
walks are discussed.

\end{abstract}
\maketitle

\medskip
\medskip

Persistence in classical dynamical systems is a topic which has been extensively studied  
during the recent years \cite{satya}.  
The persistence
probability ($P_{cl}(t)$) that the order parameter 
in a magnetic system  has not changed 
sign till time $t$ \cite{derrida} and the persistence of unvisited sites
in a diffusion problem \cite{bray} are common examples which have received   
a lot of attention.
The importance of persistence phenomena lies in the fact that the persistence
probability  
in many systems
 shows an algebraic decay (in time) with an  
exponent not related to any other known 
static or dynamical exponents.

The dynamics of a quantum system is expected to be different 
from the corresponding classical counter part, yet investigating persistence
behavior in the quantum case has remained an interesting
open problem till date.
However, defining quantum persistence is
ridden with the fundamental problem of measurement, since
in order to ensure whether or not the system persisted in a given state
(or in a chosen subspace ${\mathbf X}$), one has to impose a continuous monitoring which 
would change the dynamics of a quantum system in some essential way.
Hence in the quantum case, meaningful definition of persistence has to
include the associated measurement scheme (i.e., how the evolution
is disturbed by the measurement), and 
should essentially be dependent on that.
The dynamical process we consider here is a discrete quantum random walk (QRW) \cite{aharonov,nayak1,amban,kempe}.
Classical random walk (CRW) on a line is a much studied topic \cite{chandra,book,redner}
where at every step
one tosses a fair coin and takes a step, either to the left or right. 
The unitary implementation of QRW may be achieved through coupling an additional 
degree of freedom (a quantum coin) with the walker. This coin degree of freedom is
called the chirality, which takes values ``left" and ``right", analogous 
to Ising spin states $\pm 1$, and directs the motion of the particle. 
The state of the walker is expressed in the $|x\rangle |d\rangle$ basis,
where $|x\rangle$ is the position (in real space) eigenstate
and $|d\rangle$ is the chirality eigen state 
(either ``left'' or ``right, denoted by 
$|L\rangle$ and $|R\rangle$ respectively).
There may be several ways of choosing  the unitary operator causing 
the rotation of the chirality state, 
 conventional
choice effecting 
the rotation of the chirality state is the  Hadamard coin \cite{nayak1,amban,kempe} unitary operator.
(Most of the results are, however,  believed to be coin independent.)
The rotation is followed by a  translation represented by the operator $T$:
\begin{equation}
\begin{array}{l}
T |x\rangle |L\rangle \rightarrow   |x-1\rangle|L \rangle\\
T |x\rangle |R\rangle \rightarrow   |x+1\rangle|R \rangle\\
\end{array}.
\end{equation}
\noindent
The two component  wave function $\psi({x},t)$ describing the position of the particle is written as
\begin{equation}
\psi({x},t)=
  \begin{pmatrix}
    \psi_{L}({ x},t)\\
    \psi_{R}({ x},t) 
  \end{pmatrix},
\end{equation}
\noindent
and the occupation probability of site $x$ at time $t$ is given by
\begin{equation}
f({ x},t) = |\psi_{L}({ x},t)|^2 + |\psi_{R}({ x},t)|^2;
\end{equation}
\noindent
normalization implying $\sum_x f({ x},t) = 1$.

In the case of the  classical random walker, two well studied 
quantities are persistence or the
survival probability $P_{cl}({ x},t)$  defined as  
the probability that the  site at $x$ has not been visited till time $t$ \cite{chandra,book} and  
 the first passage time $F_{cl}({ x},t)$ which is the probability
that the walker has reached the site $x$ for the first time at time $t$ \cite{redner}.
The two
quantities are related by
\begin{equation}
 F_{cl}({ x},t)=-\frac{\partial P_{cl}({ x},t)}{\partial t}.
\label{class}
\end{equation}
At $t>>1$,   both  the persistent probability and first passage 
probabilities decay algebraically in time with exponents $\alpha_{cl}$ 
and $\beta_{cl}$ which obey the relation 
\begin{equation}
\alpha_{cl} = \beta_{cl}-1,
\label{relation}
\end{equation}
  consistent with the equality in (\ref{class}). 
We are primarily interested in the analogues of these two quantities in the QRW. 

 We now define our measurements schemes and 
the observables to quantify the concept of 
quantum persistence in this case. 
In order to measure persistence in strict sense, one 
is left with no other choice than to monitor the system continually over time.
One way of achieving  this is to impose a direct time-continual projective
measurement that determines at every moment whether or not the system persists within the 
subspace ${\mathbf X}$ in question.
In this discrete time version of quantum walk,
this amounts to carrying out a measurement after every time step of the unitary evolution 
following the scheme described below.
The walk starts from some given site at $t = 0 $, 
and a detector is placed at some other given
site $\bar x$, which detects the particle 
 with probability unity
 if it reaches there. 
If the particle is detected at $\bar x$, the evolution is stopped (here, ${\mathbf X}$ is
the entire lattice excluding $\bar x$).
Now the question  asked for such a system (rather, for an ensemble of such systems),
is what is the probability that the detector does not click till time $t$.
This is the persistence probability $P( {\bar x},t)$. 
It is equivalent to carrying out measurements at the site $\bar x$ 
after each step of unitary evolution of the ensemble 
and calculating the probability 
from the fraction of the surviving copies
(for which $\bar x$ is yet unvisited) at each step. Within this setup of QRW on a  
line, placing the detector at $\bar x$ amounts to having a semi infinite
walk (SIW) \cite{amban,absorb1,bach,Kiss} with an absorbing boundary at $\bar x$ and an open end in the other direction. 
Let us give a concrete illustration of the scheme  with a detector placed at 
$\bar x =1$. 
Suppose the walker starts at $x=0$ with left chirality.
At time $t=1$
in fifty percent cases it will be detected at $\bar x$ and the time evolution will be stopped. 
Persistence probability is therefore $1/2$ for $\bar x$ at $t=1$. 
The remaining fifty percent walks will
evolve unitarily to the next step $t=2$.  
At $t=3$, the normalized probabilities at $x=-3$, $x=-1$ and $x=1$ are 
equal to 1/4, 1/2 and 1/4 respectively (and zero elsewhere). Hence now the detector detects 
the walker at $\bar x$ with probability 
1/4, which means the 3/4 fraction of the population that was carried over to
$t = 2$ would be carried over to the next time step at $ t = 3$. This is
$3/8$ of the initial population (at $t = 0$). Hence the persistence probability
at $t = 3$ will be $3/8$ for $\bar x = 1$. 

At each time step the ensemble is measured, and
the amplitudes describe only to the surviving copies, and  the 
probabilities are to be renormalized. Let the normalized occupation probability
at  $x$ at time $t$ be 
 denoted by $\tilde f(x,t)$. Thus $\tilde f(\bar x, t^\prime)$  
 denotes the 
fraction of the copies that survived the measurement at time $t^{\prime} - 1$ (not the
fraction of the initial population) 
which reaches $\bar x$ at time $t^{\prime}$.
The persistence probability is hence given by 
\begin{equation}
P_{SIW}( {\bar x},t) =  \prod_{t^\prime=1}^{t}(1-\tilde f ( {\bar x},t^\prime)).
\label{PSIW} 
\end{equation}
It is to be mentioned here that by placing the detector at $\bar x$, it is possible to find the occupation probabilities for all $x$ and $t$ (which are strongly dependent on
$\bar x$) but the persistent probability is obtained only for $x= \bar x$. 
\noindent
One may define a first passage time $F_{SIW}({\bar x},t)$ analogous to the
classical random walk in this case as follows:
\begin{equation}
F_{SIW}({\bar x},t)=\prod_{t^{\prime}=1}^{t-1}(1- \tilde f({\bar x},t^{\prime}))\tilde f({\bar x},t)
\label{FPSIW}
\end{equation}
$$=P_{SIW}( {\bar x},t-1)\tilde f({\bar x},t).$$
It may be mentioned here that  some  related studies have been made earlier 
 \cite{absorb1,bach} and the problem of persistence measured in this 
way  had been addressed with the boundary kept 
far from the starting point of the walker \cite{bach}.

As mentioned before, definition  of quantities in a quantum system depend 
heavily on the measurement scheme and we next pose 
similar interesting and well defined questions that brings out the more intrinsic 
characteristics of the dynamics somewhat directly,  
by monitoring what we call the succession probability $S({\mathbf X},t)$ defined as follows.
Let us consider a system allowed to evolve
unitarily from a given initial state at $t_{i} = 0$, up to a 
terminating time $t^{\prime} = \Delta t$,
when finally a measurement is done on it in order to determine whether or not it resides
at a given state (or within a subspace) ${\mathbf X}$
and the evolution is stopped (e.g., in the QRW, one discards the  walk). 
 The entire process is
repeated for increasing terminating times: $t^{\prime} = \Delta t, 2\Delta t, 3\Delta t, ... t$.
Now the question is asked, what is the probability that the system will be found within ${\mathbf X}$
in every measurement with $t^{\prime} \leq t$.

 For a continuous-time evolution, this probability
will be called the succession probability
$S({\mathbf X},t)$ in the limit $\Delta t \rightarrow 0$. For a discrete random walk, 
$\Delta t$ will correspond to a single step. 
For example, in the context of QRW, one might choose to calculate the probability 
of a random walker not being found at some target site $\bar x$ in the successive
measurements done at  $t^{\prime} = 1, 2 ... t$ (starting from a given site). 
The subspace ${\mathbf X}$ in this case is constituted of all lattice points the walk may include, 
excluding $\bar x$ and we may then use the notation $S(\bar x, t)$ to denote the succession probability. It may be noted that in the classical case, 
one need not restart the evolution, after each measurement, 
since the measurement would not disturb it.
$S({\bar x},t)$ clearly differs from $P_{cl}({\bar x},t)$ in general, since
in case of $S({\bar x},t)$, in calculating the probability of finding
the system within ${{\mathbf X}}$ at $t^{\prime}$, one takes contributions
of all the paths running from  the initial time 0 
up to the final time $t^{\prime}$, including those which 
went out of ${{\mathbf X}}$ in the intermediate times. 
 
In the present one dimensional setting 
this amounts to allowing the system to evolve unitarily in either direction
(infinite walk or IW \cite{nayak1}) for an interval $t^\prime$, when the measurement is done and the
walk is discarded. We choose, again, ${\mathbf X}$ to be the entire lattice except
a given point $x = {\bar x}$, where we would like to see whether the particle has reached or not.   
To determine $S({\bar x},t)$ for a given $t$, 
the termination time $t^\prime$ is varied as  $t^\prime = 1,2 ... t$,  
and for each given $t^\prime$ one 
determines the occupation probability $f({\bar x},t^{\prime})$ and
calculates $S({\bar x},t)$ as:  
\begin{equation}
S({\bar x},t)=\prod_{t'=1}^{t}(1-f({\bar x},t^{\prime})).
\label{periw}
\end{equation}
An analogue of the the first passage time may also be
defined as 
\begin{equation}
F_{IW}({\bar x},t)=\prod_{t^{\prime}=1}^{t-1}(1-f({\bar x},t^{\prime}))f({\bar x},t)
\label{fpiw}
\end{equation}
$$=S({\bar x},t-1)f({\bar x},t).$$
Experimentally this corresponds to simply knowing 
$f({\bar x},t^\prime)$ for $t ^\prime \leq t$.\\

Some time dependent features in this type of 
infinite walk have been studied like the hitting time, recurrence time, Polya number  etc. \cite{hitting1,hitting2,Stefanak1, Stefanak2},  which involve the first passage time. However, in these studies, the spatial dependence has not been considered. For example, quantities like first passage time specifically at the origin has been estimated.

\begin{figure}
\noindent \includegraphics[clip,width= 4cm, angle=270]{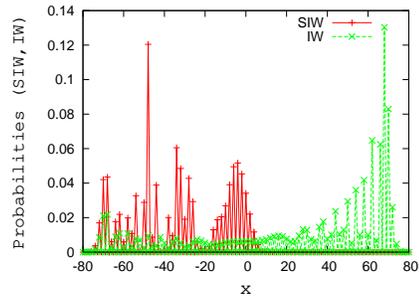}
\caption{(Color online)
 Comparison of the probabilities $\tilde f(x,t) $ for the semi infinite walk (SIW) and $f(x,t)$ 
for the infinite walk (IW) at time $t$ = 100: 
in the IW probabilities  extend to both sides, 
in the SIW (with a detector placed at ${\bar x}=10$),
the particle is not found beyond $x=10$. 
}
\label{comparison}
\end{figure}

It is important to note here that the quantities $S$ and $P_{SIW}$ 
given by equations (\ref{periw}) and (\ref{PSIW}) 
(as also $F_{IW}$  and $F_{SIW}$) 
are identical in form: the difference being 
 $ f$ appearing in the infinite walk  in place of   $\tilde  f$ in the semi
infinite walk.
Thus  $f$ and $\tilde f$  
essentially make these quantities different. As an example, we 
have shown $f(x,t)$ and $\tilde f(x,t )$ as functions of $x$ at a fixed time $t$ in Fig \ref{comparison}.
For the semi infinite walk, there is a detector placed at $x=10$. 
To emphasize the difference, we have generated a walk biased towards the 
right, and the unbounded walk shows it clearly. On the other hand, in the 
semi infinite walk, the walker is not allowed beyond $x=10$ and consequently is
 driven towards the left. Obviously, even if a walk with symmetric boundary condition is initiated, the 
presence of the detector will convert  it to a asymmetric walk.

In the calculation, 
 a quantum random walk is initialized at the origin with
$\psi_{L}(0,0) = a_0,  \psi_{R}(0,0)= b_0; ~~a_0^2 + b_0^2 =1.$
(All other $\psi_{L}$  and  $\psi_{R}$ taken equal to zero).
 $\psi_{L}({\bar x},t)$ and $\psi_{R}({\bar x},t)$ are recursively evaluated for all $x$ and $t$. In the bounded (semi infinite) walk, contributions from the walks
going through $\bar x$ are ignored.
Unless otherwise specified, we have taken  
 $a_0=1/\sqrt{2}, b_0=i/\sqrt{2} $ which would result in a symmetric 
walk for the unbounded (infinite) walk case.

\begin{center}
\begin{figure}
\vskip -2cm
\hskip 1cm
\noindent \includegraphics[clip,width= 10cm]{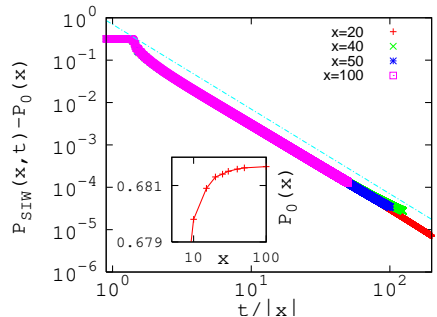}
\caption{(Color online) The data collapse for the residual persistence probability   in the 
 quantum random walk shown for different values of $x=\bar x$. 
The straight line in the log-log plot has the slope = -2.0.
Inset shows the variation of the saturation value $P_0$ with $x$.}
\label{collapse2}
\end{figure}
\end{center}

\begin{figure}
\noindent \includegraphics[clip,width= 4cm,angle=270]{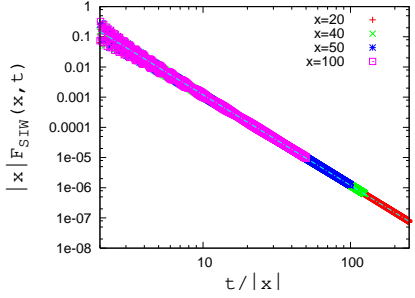}
\caption{(Color online) The data collapse for the first passage  probability in the  
 quantum random walk shown for different values of $x=\bar {x}$.
Fitted straight line has slope = -3.0.
}
\label{collapse3}
\end{figure}

The results for SIW are essentially numerical;
 the persistence probabilities
here   saturate  in time.
This saturation behavior apparently originates from the simultaneous effect of 
drifting of the quantum walker away from the origin and the presence of the 
boundary at $\bar x$.
These observations are in  agreement with \cite{bach} and consistent with other results
involving recurrence time etc. \cite{amban,Kiss}.
The first passage times, on the other hand, decay algebraically with $t$.
As already mentioned, in \cite{bach}, the persistence probability for large $\bar {x} $ was found to vary as  
\begin{equation}
P_{SIW}({\bar x},t) =  P_0 + {\rm{const} } (t/{|\bar {x}|})^{-\alpha_{SIW}}
\label{psiw}
\end{equation}
where $P_0$ is the saturation value and  
$\alpha_{SIW}=2$. We verify this result with the observation that 
$P_0$ has a weak dependence on $\bar {x}$ and observe that  the  numerical value of 
$\alpha_{SIW} $ 
approaches  the value 2 asymptotically. 
The numerically estimated values of  $P_0(\bar {x})$ are found to vary   as $(a-b\exp(-c\bar {x}))$ with
$c =0.30 \pm 0.02$, shown in the inset of Fig \ref{collapse2}.   
Using these  values of $P_0(\bar {x})$, we show that a data collapse is obtained when the
residual persistence probability $P_{SIW}({\bar x},t) -  P_0(\bar x)$ 
is plotted against $t/|\bar{x}|$.   
The 
first passage time 
 $F_{SIW}({\bar x},t)$ behaves as
\begin{equation}
F_{SIW}({\bar x},t) \propto  (t/|{\bar x}|)^{-\beta_{SIW}}/|{\bar x}|,
\label{fsiw}
\end{equation}
with  $\beta _{SIW}\simeq 3.0$.
Results for the collapsed data of persistence and first passage times are
shown in Figs \ref{collapse2} and \ref{collapse3}.


For the unbounded or infinite walk, the calculations can also be done using the
analytical forms  available in \cite{nayak1}
\begin{equation}
 \psi_{L}({ x},t)=\frac{1+(-1)^{x+t}}{2}\int\frac{dk}{2\pi}(1+\frac{\cos k}{\sqrt{1+\cos^{2}k}})e^{-i(\omega_{k}t+kx)}
\label{analytic1}
\end{equation}
\begin{equation}
 \psi_{R}({ x},t)=\frac{1+(-1)^{x+t}}{2}\int\frac{dk}{2\pi}\frac{e^{ik}}{\sqrt{1+\cos^{2}k}}e^{-i(\omega_{k}t+kx)}
\label{analytic2}
\end{equation}
(which are obtained for a initial state with left chirality, i.e., $a_0=1, b_0=0$)
and  evaluate $f({\bar x},t)$ directly by numerical integration.

\begin{figure}
\noindent \includegraphics[clip,width= 4cm, angle=270]{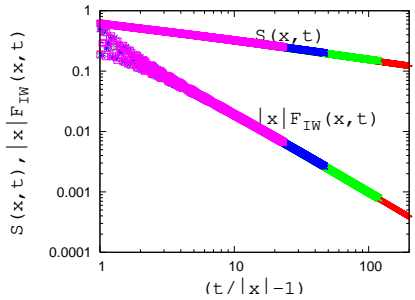}
\caption{(Color online) The data collapse for the succession  probability and scaled 
first passage probability defined for a infinite quantum random walk shown for different values of $x =10$ (red), $20$ (green), $50$ (blue), $100$ (magenta). }
\label{collapse1}
\end{figure}

Power law decay for both 
  $S({\bar x},t)$  and $F_{IW}({\bar x},t)$  are observed:
\begin{equation}
S({\bar x},t) \propto (t/|{\bar x}|-1)^{-\alpha_{IW}}
\label{perseq}
\end{equation}
for $t/|{\bar x}| > 1$,  
and 
\begin{equation}
F_{IW}({\bar x},t) \propto (t/|{\bar x}|-1)^{-\beta_{IW}} /|{\bar x}|
\label{firsteq}
\end{equation}
for  $t/|{\bar x}| >> 1$  with 
$\alpha_{IW} \simeq 0.31$ and $\beta_{IW} \simeq 1.31$. 
Data collapse  for  $S({\bar x},t)$  and $F_{IW}({\bar x},t)$ from the numerical evolution of the infinite walk are shown in Fig. {\ref{collapse1}}. 
These  results are obtained with 
  $a_0=1/\sqrt{2}, b_0=i/\sqrt{2} $ which correspond to  a symmetric walk.
Results obtained from the numerical integration of Eqs. \ref{analytic1} 
and  \ref{analytic2} (corresponding to $a_0=1, b_0=0$ giving an asymmetric walk) show the 
same scaling behaviour as $S({\bar x},t)$  and $F_{IW}({\bar x},t)$. 
Thus the exponents are independent of the initial conditions as expected.

As discussed in \cite{Stefanak1,Stefanak2},
 the quantum walk on a line is recurrent, i.e. it returns to
the origin with certainty and the same applies to visiting any other
lattice point. Hence, the  asymptotic succession probability is zero, which is in
agreement with the power law decay of $S(\bar x,t)$ found in the present paper.

\begin{figure}
\noindent \includegraphics[clip,width= 4.cm, angle=270]{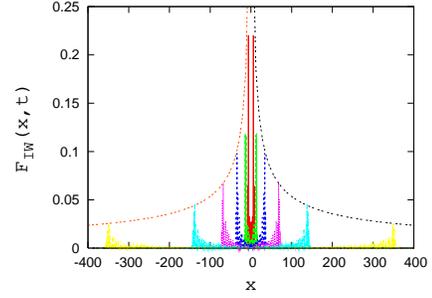}
\caption {Typical variation of $F_{IW}(x,t)$ against $x$ for
different values of $t=10,20,50,100,200,500$. The curves extend over larger 
values of $x$ as $t$ is increased. The peaks are approximately fitted to $0.81x_{max}
^{-0.59}$.}
\label{fmax_x}
\end{figure}

\begin{figure}
\noindent \includegraphics[clip,width= 4.cm, angle=270]{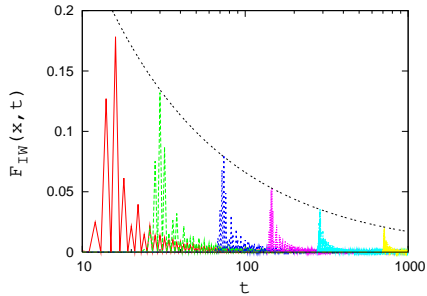}
\caption{Typical variation of $F_{IW}(x,t)$ against $t$ for different values of $x= 10,20,50,100,200,500$.  
The curves extend over larger 
values of $t$ as $x$ is increased.
The peaks are approximately fitted to $t_{max}^{-0.59}$.}
\label{fmax_t}
\end{figure}

While the exponents $\alpha_{IW}$ ($\alpha_{SIW}$) and $\beta_{IW}$ ($\beta_{SIW}$) 
are different
from the classical $\alpha_{cl}$ and $\beta_{cl}$, they enjoy a relationship
identical to eq (\ref{relation}). We consider some other quantities related to 
the function  $F_{IW}(x,t)$ which one can  compare   with their classical 
counterparts. Plotting $F_{IW}(x,t)$ against $x$ or
$t$, we notice that it has an oscillatory behavior. These oscillations which die down for large values
of $t/x$ as is apparent from Fig. \ref{collapse1} can be traced to
the oscillatory behavior of $f(x,t)$ for a QRW observed earlier  \cite{nayak1,amban,kempe}.
 From Figs \ref{fmax_x} and \ref{fmax_t}, we observe that $F_{IW}(x,t)$ actually attains a
maximum value $F_{IW,max}(x,t)$ at values of $|x|=x_{max}$ (or $t=t_{max}$) for fixed values of $t$ (or $x$).
We notice that
$F_{IW,max}(x_{max},t) \propto x_{max}^{-\delta}$ where $\delta \simeq 0.59$.
Keeping $x$ fixed,  $F_{IW,max}(x,t_{max})$ versus
$t_{max}$ shows the  same kind of dependence, i.e., $F_{IW,max}(x,t_{max}) \propto t_{max}^{-\delta}$. That the scalings with $t_{max}$ and $x_{max}$ turn out to be identical  is
not surprising as $x$ scales as  $t$ in a QRW.
 It is not possible to obtain this scaling form directly from eq. (\ref{firsteq})   since
 $F(x,t)$ attains a
maximum value when  $t/|x|$ is close to unity where the   fitted scaling form is not exactly  valid.
In fact, eq.
(\ref{firsteq})  does not give any maximum value at all.

Another dynamic quantity called  hitting time has been estimated earlier for the QRW,
in which an  absorber is assumed to be located at a  specific vertex of a hypercube within which the
walk is conceived \cite{hitting1,hitting2}. The average hitting time is by definition the average time to reach that
particular vertex for the first time.
One can  evaluate the
average hitting time $\tau_h(x)$ in the infinite walk scheme using  $\tau_h(x) = \sum_0^T t F_{IW}(x,t)$
where $t$ is allowed to vary from $0$ to $T$:
\[
\tau_h \approx \int_0^{T}t F_{IW}(x,t)dt
\]
\[
\sim {T}^{2-\beta_{IW}} x^{\beta_{IW} -1}/(2-\beta_{IW}) + O({T}^{-{\beta_{IW}+1}}).
\]
The numerical data (not shown) gives a fairly good agreement with this scaling.
The above equation shows that  $\tau_h$ blows up for $T \to \infty$ in agreement with some earlier results using other
coins \cite{hitting1,hitting2}.

We thus observe that a  number of quantities related to the dynamics 
of a quantum random walker follow power law behavior with time. Of these,
the persistence probability $P_{SIW}(\bar x, t)$, which is obtained from a semi infinite walk, is drastically different
from its classical analogue as it approaches a constant value in a 
power law fashion with
an exponent which is quite different from the classical value 1/2.
The first passage time also has a power law decay with a new exponent.
The numerical data also indicate that the two quantities  
obey a simple relation as in 
 the classical case (eq (\ref{class})). 

A different quantum measure which we call the  succession probability has  been 
proposed and calculated in the present work and a corresponding first passage
probability defined. These measures can be obtained from an infinite
 walk. The persistence probability and succession probabilities are
similar in form but the results are highly different with the succession probability exhibiting a power 
law decay  (no saturation) with yet another 
new exponent.  
However, the form of the  probabilities in eqs (\ref{psiw}), (\ref{fsiw}) and 
(\ref{perseq}),
(\ref{firsteq}) and the values of the exponents indicate the validity of eq (\ref{class}) in the quantum case as well.

In a classical random walk,  $\langle x^2\rangle \propto t^{\gamma_{cl}}$ with $\gamma_{cl}=1$ and in one dimension this scaling 
  governs all other dynamic behavior including persistence. 
Thus all other exponents  like $\alpha_{cl} $ and $\beta_{cl}$ are essentially 
dependent on $\gamma$, e.g., $\alpha_{cl} = \gamma_{cl}/2$ and $\beta_{cl} = \frac{3}{2}\gamma_{cl}$. 
 A quantum walker propagates much faster; here $\langle x^2\rangle \propto t^{\gamma_{q}}$ with  $\gamma_q=2$.     
Thus the   dimensionless factor $x/t$ appears in the scaling argument of 
the dynamic quantities. 
The exponents $\alpha_{SIW}\simeq 2 $ and $\beta_{SIW} \simeq 3$ appear to be simply  related to
 $\gamma_q$; $\alpha_{SIW} = \gamma_q$ and $\beta_{SIW} = \frac{3}{2} \gamma_q$ showing 
that here too the persistence phenomena is governed by the scaling 
 $\langle x^2\rangle \propto t^{\gamma_{q}}$
 only.

    
For the infinite walk case,  the exponents $\alpha_{IW}$ and $\beta_{IW}$ are
  apparently not simply related to $\gamma_q$.
We have estimated  some additional quantities involving the 
first passage time.
The     maximum values of the classical probability $F_{cl}$,
 behaves as $1/t_{max}$ (for $x$ constant) or $1/x_{max}^2$ (for
$t$ constant) showing that
the obtained exponents are
simple multiples of $\gamma_{cl} =1$.
On the other hand,
the behavior of $F_{IW,max}$  appears to
 depend on the
value of $\alpha_{IW}$  and not $\gamma_q$ as it varies with  ${t_{max}}$ or
$x_{max}$ with an exponent  $\delta$ which is very close to
$2\alpha_{IW}$ numerically.

The average hitting time for
a CRW  is found to vary as $T^{\gamma_{cl}/2}$.
In the infinite QRW, this variation is given by $T^{2-\beta_{IW}}$. For the classical case,
$2-\beta_{cl} = \gamma_{cl}/2$  but since no such relation exists for the
quantum case, the hitting time scaling is therefore {\it{not}} dictated by
$\gamma_q$ but by  $\beta_{IW}$ (or $\alpha_{IW}$) only.

Lastly, it is true that the probabilities $\tilde f({\bar x},t)$ 
are quite different from $f({\bar x},t)$ making the persistence and succession probabilities 
distinct,  however, the 
feature that the quantum walker walks away from the origin (in contrast to a classical walker) is present in both.
This makes the persistence probability and the succession probabilities quite large in magnitude compared
to classical persistence probabilities. For the quantum persistence probability,
$P_{SIW}$,
the additional constraint of the presence of the boundary makes it saturate.

Acknowledgment: Financial supports from DST grant no. SR-S2/CMP-56/2007 (PS)  and 
UGC sanction no. UGC/209/JRF(RFSMS) (SG) are acknowledged.


\begin{thebibliography}{99}

\vskip -5cm

\bibitem{satya} S.N. Majumdar, 
Curr. Sci. {\bf{77}},  370 (1999). 

\bibitem{derrida} B. Derrida, A.J.Bray and C. Godreche, J.Phys. A {\bf {27}}, L357  (1994).
\bibitem{bray} S. J. O'Donoghue and A. J.  Bray, Phys. Rev. E {{\bf 64}}, 041105 (2001);
P. K. Das, S. Dasgupta and P. Sen, J. Phys. A: Math. Theor. {\bf {40}}, 6013 (2007). 
\bibitem{chandra} S. Chandrasekhar, Phys. Rev. {\bf {15}}, 1 (1943). 
\bibitem{book} G. H. Weiss, { {Aspects and applications of the random walk}}, North Holland, Amsterdam, 1994.
\bibitem{redner} S. Redner,  {{A guide to first-passage processes}}, Cambridge University Press, 2001.
\bibitem{aharonov} Y. Aharonov, L. Davidovich and N. Zagury, Phys. Rev. A {\bf {48}}, 1687 (1993).  
\bibitem{nayak1} A. Nayak and A. Vishwanath, DIAMCS Technical Report 2000-43 and Los Alamos preprint archive,quant-ph/0010117.
\bibitem{amban} A. Ambainis, E. Bach, A. Nayak, A. Vishwanath and J. Watrous, Proc. 33th New York, NY (2001).
\bibitem{kempe} J. Kempe, Contemporary Physics {\bf {44}}, 307  (2003).
\bibitem{absorb1} T. Yamasaki, H. Kobayashi, and H. Imai, 
     arxiv: quant-ph/0205045.
\bibitem{bach} E. Bach, S. Coppersmith, M. Paz Goldshen, R. Joynt and J. Watrous,
 Journal of Computer and Systems Sciences, {\bf 69}, 562
 (2004).
\bibitem{Kiss}T. Kiss, L. Kecskes, M. Stefanak, I. Jex, Physica Scripta {\bf T135}, 014055 (2009). 
\bibitem{hitting1} 
J. Kempe, Proc. 7th RANDOM,  354 (2003); J. Kempe,  Probability Theory and Related Fields, {\bf 133}, 215  (2005).
\bibitem{hitting2} H. Krovi and T.A. Brun,  Phys. Rev. A {\bf 73}, 032341 (2006). 
\bibitem{Stefanak1} M. Stefanak, I. Jex, T. Kiss, Phys. Rev. Lett {\bf 100}, 020501 (2008).
\bibitem{Stefanak2}
 M. Stefanak, T. Kiss and I. Jex,  Phys. Rev. A {\bf 78}, 032306 (2008).
\end{thebibliography}
\end{document}